# Mapping Research Data at the University of Bologna

Chiara Basalti[1], Giulia Caldoni[1*], Sara Coppini[1], Bianca Gualandi[1,2], Mario Marino[1], Francesca Masini[3] and Silvio Peroni[2]

[1] Research Division (ARIC), Alma Mater Studiorum - Università di Bologna, Bologna, Italy

[2] Department of Classical Philosophy and Italian Studies, Alma Mater Studiorum - Università di Bologna, Italy

[3] Department of Modern Languages, Literatures, and Cultures, Alma Mater Studiorum - Università di Bologna, Bologna, Italy

* Corresponding author: giulia.caldoni2@unibo.it

## Author's Contributions

**Chiara Basalti**: conceptualization, project administration, supervision

**Giulia Caldoni**: conceptualization, data curation, methodology, writing – original draft, writing – review & editing

**Sara Coppini**: conceptualization, data curation, methodology, writing – original draft

**Bianca Gualandi**: conceptualization, data curation, methodology, writing – original draft, writing – review & editing

**Mario Marino**: conceptualization, software, data curation, methodology, writing – original draft, writing – review & editing

**Francesca Masini**: conceptualization, project administration, supervision, writing – review & editing

**Silvio Peroni**: conceptualization, project administration, supervision, writing – review & editing

## Abstract

This paper presents an analysis of research data management (RDM) practices at the University of Bologna (UniBO). It is based on qualitative data manually collected from 29 data management plans (DMPs) and recorded within a table based on 27 relevant 'DMP variables'. The study examines aspects such as data types and formats, handling of personal data, and repository selection for long-term preservation.

Using descriptive statistics, we address three core questions:

1. What is the diversity of research data managed at UniBO?
2. What recurring challenges and patterns can inform data stewardship services?
3. Is there evidence of interdisciplinary data production?

Findings reveal significant overlaps across disciplines in terms of data types and formats, with limited emphasis on data reuse, while common challenges include confidentiality, privacy protection, intellectual property rights, and long-term preservation.

The results underscore growing awareness of RDM and provide insight on the type of institutional training and support that should be prioritized to address emerging needs.

## Introduction

Research data management (RDM) has become increasingly relevant due to the growing adoption of Open Science policies and practices. As stated in UNESCO's Recommendation on Open Science (UNESCO, 2021), Open Science introduces a paradigm centered on reproducibility, transparency, sharing, and collaboration, promoting broader access to scientific outputs such as publications, data, software, hardware, and educational resources. To support this, research outputs must adhere to principles of reliability, transparency, and reproducibility (Leonelli, 2018; Credi, Masini, *et al.*, 2024).

Borghi and Van Gulick (Borghi and Van Gulick, 2022) highlight the overlap between RDM, reproducibility, and Open Science, noting that each offers a lens to interpret the other. Trustworthy data sharing depends on well-structured, documented, and licensed datasets, managed according to FAIR principles (Wilkinson *et al.*, 2016) and deposited in repositories for long-term preservation. The principle 'as open as possible, as closed as necessary' (EC, no date) allows for justified restrictions to ensure data security, privacy, consent compliance, and cost management (NIH, 2020; DG for Research and Innovation (EC), 2021).

Policy-level support for RDM is growing (DG for Research and Innovation (EC), 2021; NIH, 2020), yet researchers often are forced to acquire new skills (Borghi and Van Gulick, 2022), which are rarely formally acknowledged (Fecher, Friesike and Hebing, 2015). Disciplinary differences in data types and standards contribute to a fragmented RDM landscape (Akers and Doty, 2013; Gualandi, Pareschi and Peroni, 2023).

If discipline-specific approaches become especially relevant in interdisciplinary research, where integrating theories, concepts, techniques, and data from multiple fields is essential (Stokols *et al.*, 2003; Schröder *et al.*, 2020), shared language and semantic interoperability are crucial for effective data exchange (Park, 2022; RDA, no date). And while collaboration is widely recognized as a driver of scientific progress (Li, Chen and Larivière, 2023), evaluating interdisciplinarity remains an open challenge (Chen *et al.*, 2024). .

Many Research Performing Organizations (RPOs) have developed strategies to promote RDM and Open Science, including institutional policies and data stewardship services aligned with the Strategic Research and Innovation Agenda (SRIA) of the European Open Science Cloud (DG for Research and Innovation (EC) and EOSC Executive Board, 2022). Data stewards are key professionals in fostering a culture of data sharing and supporting researchers throughout the data lifecycle (Gualandi, Caldoni and Marino, 2022; Basalti *et al.*, 2024), with a focus on both new data and reuse (Borgman and Groth, 2024; Puebla and Lowenberg, 2024).

Review studies (Perrier *et al.*, 2017; Perrier, Blondal and MacDonald, 2020) show that assessing actual RDM practices can help identify challenges and suitable workflows, shifting the focus from abstract principles to practical implementation. Data stewards contribute to this shift by providing insights that implement the G7 Open Science Working Group's recommendation to base Open Science policies on research evidence (ECCC - Environment and Climate Change Canada *et al.*, 2023).

The University of Bologna (UniBO) has established a multidisciplinary team of data stewards to support researchers in implementing RDM practices and drafting data management plans (DMPs). This tailored support focuses on identifying the types of data produced or reused in each project and defining appropriate RDM strategies, which are then consolidated in the DMP.

This study, carried out by UniBO's four data stewards, aims to map and categorize institutional RDM practices, such as the choice of data types and formats, personal data handling, and repository selection, to identify disciplinary differences, shared practices, and challenges for institutional support services and infrastructures (Parland-von Essen *et al.*, 2018).

After selecting relevant DMP variables (see Table 1), their values were manually collected through close reading of the DMPs (see Materials and Methods). The analysis addresses three research questions:

1. How diverse is the range of data managed at UniBO? This explores the types, formats, and size of reused and/or generated data across disciplines.
2. What recurring issues and patterns in data management could inform or improve data stewardship services? This focuses on cross-cutting challenges that often require expert support.
3. Is there an interdisciplinary approach to data production at UniBO? This investigates cross-domain integration using the Italian national system of scientific-disciplinary sectors (SSD).

The article is structured as follows: Section 'Materials and Methods' outlines the methodology; Section 'Results' presents and discusses the findings; Section 'Discussion and Conclusions' offers final reflections and future directions.

## Materials and Methods

To conduct this study, the four data stewards analyzed 29 DMPs drafted by UniBO researchers involved in EU-funded projects (Horizon 2020 or Horizon Europe). DMPs are mandatory deliverables to be submitted within the first six months of such projects, with updates occasionally requested or recommended at a later date. Of the 29 DMPs in this sample, only four were updates of initial DMPs and reflected a more advanced stage of the respective projects.

All 29 DMPs were written between May 2022 and October 2023 by UniBO researchers, in their role as project leaders or partners responsible for the DMP, with the support of the same team of data stewards that conducted the analysis. Access to the documents, including some classified as sensitive and not publicly available, was possible thanks to this support. To ensure confidentiality, all collected information was anonymized (Coppini *et al.*, 2023, 2024).

For this study, the data stewards manually reviewed each DMP to collect information on RDM practices (e.g., data types and formats choice, personal data handling strategies, repository selection). The information collected during the manual DMP analysis was systematically recorded in a spreadsheet structured around the 27 DMP variables listed in Table 1.

These variables were derived from a preliminary analysis conducted collaboratively by the four data stewards, who examined the DMPs and other project-related documentation to identify the most pertinent dimensions for assessing institutional RDM practices.

Following consensus on the list of DMP variables and the structure of the corresponding data collection table, the data steward proceeded to analyze the DMPs. Each was assessed by a single data steward within their disciplinary expertise, with complex cases reviewed collectively. The analysis was based solely on their interpretation, no generative AI tools were used, and no verification against real-world repositories was performed.

The use of a systematically structured data collection table, alongside the implementation of controlled vocabularies for many of the recorded values, contributed to ensuring consistency across evaluations. Nonetheless, the present analysis remains inherently shaped by the individual interpretations of each evaluator.

*Table 1: List of DMP variables for data analysis with description and accepted values. (ND = Not Determined)*

| N | Field | Description and accepted values |
|---|---|---|
| 1 | Project identifier | Alphanumeric string that identifies the project. Three-digit sequential numbering, independent of other identifiers. |
| 2 | Dataset identifier | Alphanumeric string that identifies the dataset. Three-digit sequential numbering, independent of other identifiers. |
| 3 | Entry identifier | Alphanumeric string that identifies the data category (e.g., file), described in the current row. Three-digit sequential numbering, independent of other identifiers. |
| 4 | Creator's unit | Research unit (e.g., department, center) of the individual who created, reused, or contributed to the dataset. Accepted values include 'ND' (new data with undefined creator), and 'EXT' (external data created by individuals outside UniBO). Multiple values may be listed when contributors belong to different known or unknown research units. |
| 5 | Creator's SSD | Disciplinary scientific sector (SSD) of the individual who created, reused, or contributed to the dataset. Accepted values include 'ND' (new data with undefined creator), and 'EXT' (external data created by individuals outside UniBO). Multiple values may be listed when contributors belong to different known or unknown research units. |
| 6 | Principal Investigator's SSD | Disciplinary scientific sector (SSD) of the principal investigator of the project. |
| 7 | Project unit | Research unit (department, center, etc.) of the principal investigator of the project. |
| 8 | Project program | The program under which the project is being funded. Values are: HE (Horizon Europe); H2020 (Horizon 2020). |
| 9 | Project type | Classified according to the recipient of the funding. Values are: Individual; consortium. |
| 10 | Subject area | Disciplinary or thematic area to which the project belongs. |
| 11 | Month DMP is delivered | Deadline for the delivery of the DMP, expressed in the form: M6 (within six months from the start of the project), M12 (within twelve months from the start of the project). |
| 12 | Public DMP | Is the DMP publicly available? Values are: 1 (True), 0 (False). ND is also accepted. |
| 13 | Data type | Typology of data on a formal level, e.g. image. ND is also accepted. |
| 14 | Data content | Categorization of the data at the content level, e.g., scanned image of a medieval manuscript. |
| 15 | Format | Format and extension (if more than one, separated by commas). ND is also accepted. |
| 16 | New data | Is the data newly produced? Values are: 1 (True), 0 (False). ND is also accepted. |
| 17 | Contains personal data | Does the research data contain personal data? Values are: 1 (True), 0 (False). ND is also accepted. |
| 18 | Personal data management strategy | Strategy adopted to respect participants' privacy with respect to their personal data. Values are: Anonymization, pseudonymization, no strategy, consent to publish. ND is also accepted. |
| 19 | Level of access | Accessibility of the data entry. Values are: Open, accessible under conditions, controlled, embargoed, unlicensed, unfiled, unknown. |
| 20 | Reason of inaccessibility | Excessive size, ethical issues, privacy, IPR. ND is also accepted. |
| 21 | Size | Orders of magnitude for digital data (Bytes, KB, MB, GB, TB, PB, EB, ZB, YB). ND is also accepted. |
| 22 | Standard | Name of standards used to organize and structure data (e.g., vocabularies, ontologies, taxonomies). ND is also accepted. |
| 23 | Deposited | Has the data been deposited in a data repository? Values are: 1 (True), 0 (False). |
| 24 | Chosen repository | Name of repository chosen by researchers to deposit data, as stated in re3data.org. ND is also accepted. |
| 25 | PID | Does the data have an associated PID? Values are: 1 (True), 0 (False). |
| 26 | Associated publication | Does the data have an associated publication? Values are: 1 (True), 0 (False). ND is also accepted. |
| 27 | Notes | General notes concerning other unclassified issues. |

Only digital research outputs distinct from traditional publications (e.g., journal articles, books, conference papers) were considered, including both newly generated and reused data mentioned in the DMPs.

Different data types may coexist within a single dataset: for example, an interview dataset may include an audio recording and a transcript. To account for differences in format, volume, and RDM challenges, these components were treated as separate data entries. Thus, the unit of analysis is not the dataset but the individual *data entry*, defined as a single type of data within a dataset.

Existing taxonomies were used whenever possible to define field values. These included general frameworks such as DataCite for data types (DataCite Metadata Working Group, 2019) and the Italian SSD system for disciplines (LEGGE 19 novembre 1990, n. 341, no date), as well as UniBO-specific classifications (e.g., the five main subject areas of academic research). In cases where no suitable taxonomy existed new ones were developed.

Some existing taxonomies were adapted. For instance, the data stewards reused DataCite's controlled vocabulary for the field data type, selecting values from element 10.a resourceTypeGeneral: *Audiovisual*, *ComputationalNotebook*, *Image*, *InteractiveResource*, *Report*, *Software*, *Sound*, *Standard*, *Text*, *Workflow*, *Other*, *Model*. However, the term *Dataset* was renamed *Tabular* to avoid confusion, as in UniBO's context dataset refers to a collection of data, whereas in DataCite it typically denotes structured/tabular data (DataCite Metadata Working Group, 2019).

The analysis primarily employed descriptive statistics to explore the three research questions, based on the information collected in the spreadsheet structured around the defined DMP variables. The computational tool used was R (version 4.2.2).

Further details on the methodology and its implementation in R are available in Coppini *et al.* (2023) and Marino *et al.* (2024).

## Results

This section is structured around the three research questions and presents and discusses the research results.

### Research question 1. How diverse is the range of data managed at the University of Bologna?

The first research question explores the types and formats of data across disciplines, and whether they are reused or created anew.

#### 1.1 What types of data do researchers manage at the University of Bologna?

As far as data types are concerned, the data stewards looked at:

- The most popular data types in general terms,
- How often different data types are found within the same dataset,
- How often different data types are found within the same project,
- How data types are distributed across single-beneficiary and collaborative EU-funded projects,
- How data types are distributed across subject areas.

Figure 1 shows all data entries by data type. The values are taken from the DataCite taxonomy (DataCite Metadata Working Group, 2019), with one modification (see Materials and Methods).

*Figure 1 Most popular data types per data entry.*

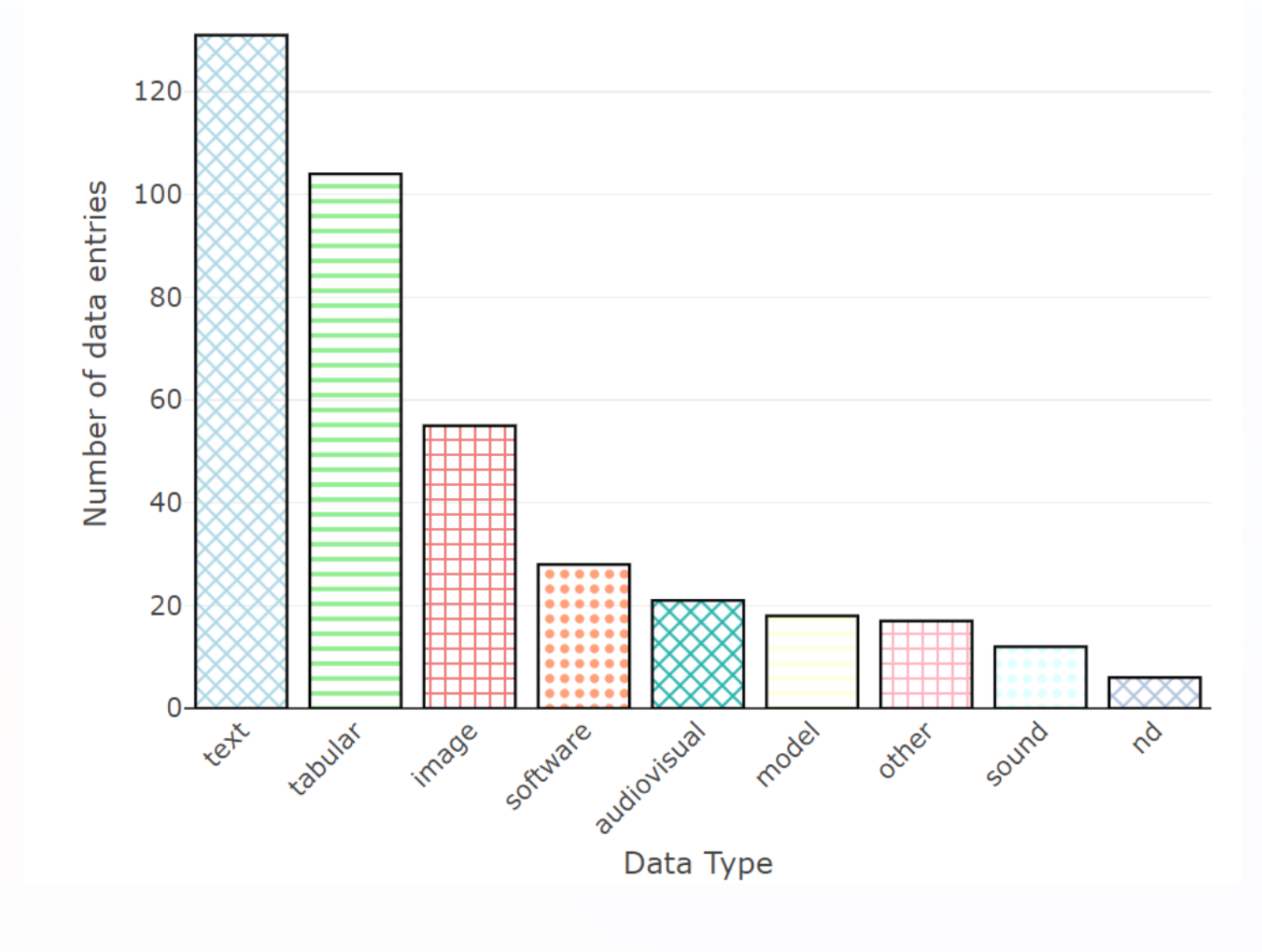

Each data entry was assigned a data type based on the description provided by researchers in their DMPs and the indicated file format. For instance, .txt, .pdf, and .rtf files were classified as *Text*; .csv, .xlsx, and .ods as *Tabular*; .jpg, .png, and .tif as *Image*, and so on. In one case, a research group used Jamboards to collect information during meetings. Although the format was .pdf, the data stewards opted for *Other* instead (see also below).

*Text* emerges as the most frequent data type (131 entries), followed by *Tabular* data (104), *Images* (55), *Software* (28), *Audiovisual* data (21), *Models* (18), and *Sound* (12). At this stage, the analysis includes both new and reused data without distinction.

In 17 cases, the category *Other* was used when the data could not be described using more specific DataCite taxonomy terms. These included experimental measurements (e.g., .bag, .m, .mat), metadata and databases (e.g., .json, .rdf), and shapefiles (e.g., .shp). In six instances, classification was unclear due to vague descriptions or unspecified formats.

Following this general overview, the data stewards examined each dataset individually. A *dataset* is defined as a collection of data intended for repository deposit, composed of one or more data entries.

A small majority of datasets (121) contain a single data type. The remaining 85 datasets include multiple types, typically between two and four (Figure 2). Researchers tend to group data by type, but it is important to note that nearly all DMPs analyzed (except four) are initial versions compiled within the first six months of the projects. These reflect intentions rather than completed actions, and bundling by type may be more common at early stages. It remains unclear whether this tendency evolves over time.

*Figure 2 Number of unique data types per dataset.*

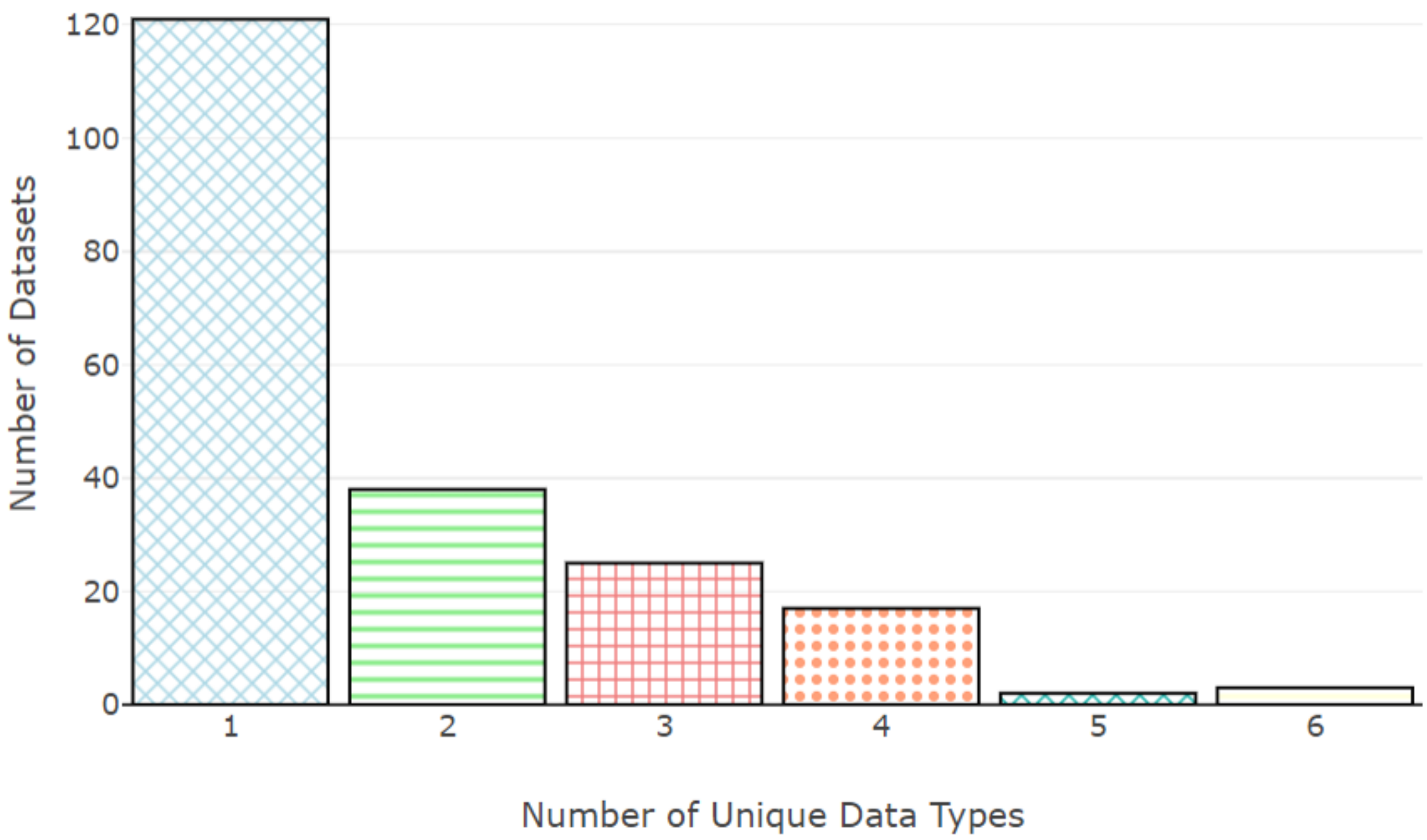

Additionally, the prevalence of single-type datasets may be overstated. Some DMPs did not specify the number of datasets to be produced, instead loosely categorizing data by type. In such cases, the data stewards grouped all data types into a single dataset to avoid artificially inflating dataset counts, though researchers may later separate them.

When grouping data entries by project, most projects handle two to four distinct data types. Projects generally produce or reuse multiple types of data, though typically fewer than five, even in large, multi-partner initiatives (Figure 3).

*Figure 3 Number of unique data types per project.*

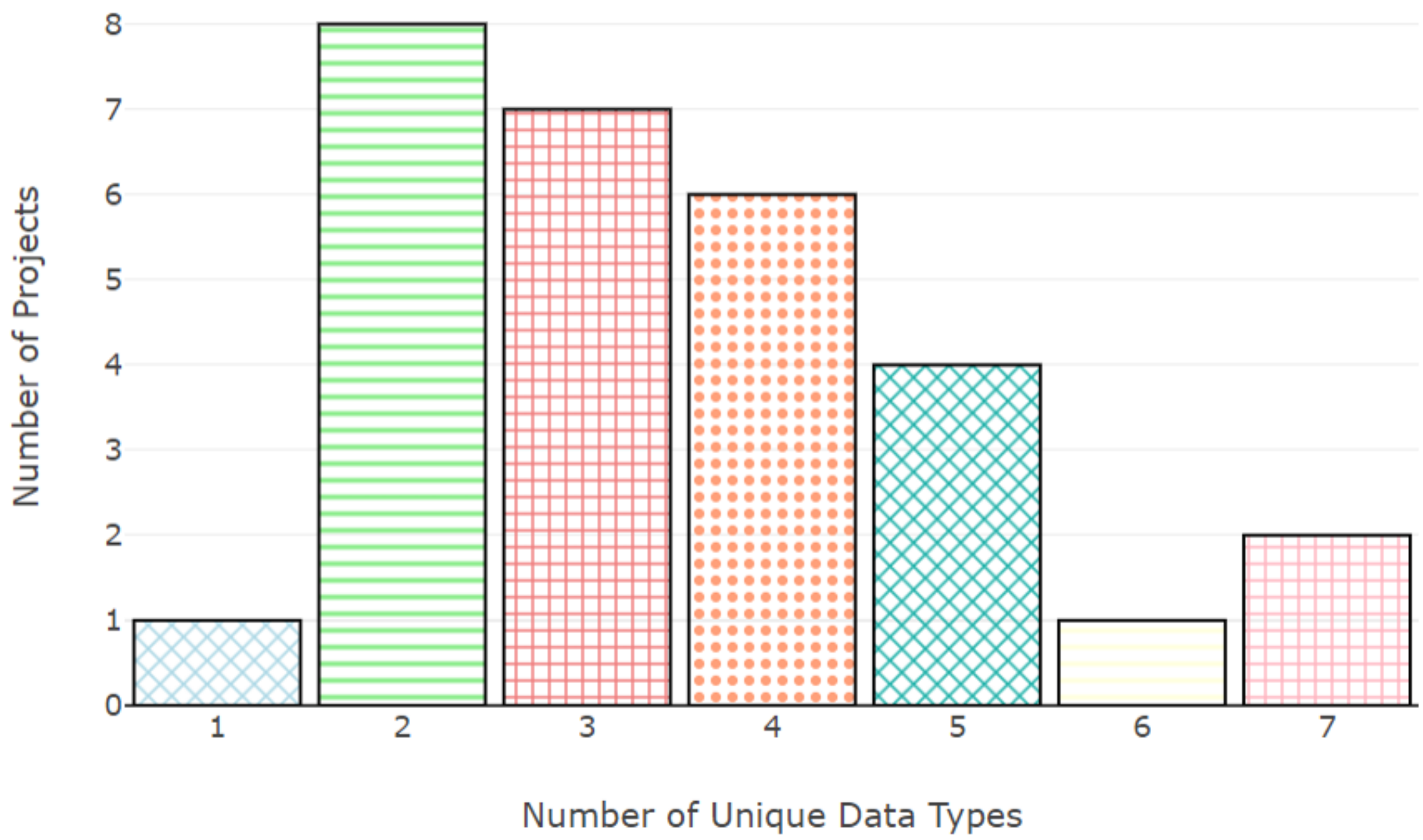

The data stewards also examined differences between single-beneficiary projects (e.g., ERC Starting Grants) and collaborative projects, with consortia often coordinated by UniBO. Of the 29 projects analyzed, 17 were single-beneficiary and 12 collaborative (see Table 2).

*Table 2: Number of single-beneficiary (tot. 17) and collaborative projects (tot. 12) by number of data types handled.*

| N of data types handled | N of single-beneficiary projects handling N of data types | N of collaborative projects handling N of data types |
|---|---|---|
| 1 | 1 | 0 |
| 2 | 3 | 5 |
| 3 | 7 | 0 |
| 4 | 2 | 4 |
| 5 | 3 | 1 |
| 6 | 0 | 1 |
| 7 | 1 | 1 |

Just over 30% of single-beneficiary projects (six out of 17) produce or reuse four or more data types. In contrast, more than 50% of collaborative projects (seven out of 12) meet this threshold. Collaborative projects tend to involve a wider variety of data types, likely due to the greater number of researchers, areas of expertise, and research strands involved (Figure 4a and 4b).

*Figure 4 Breakdown of data types chosen across (a) single beneficiary projects and (b) collaborative projects.*

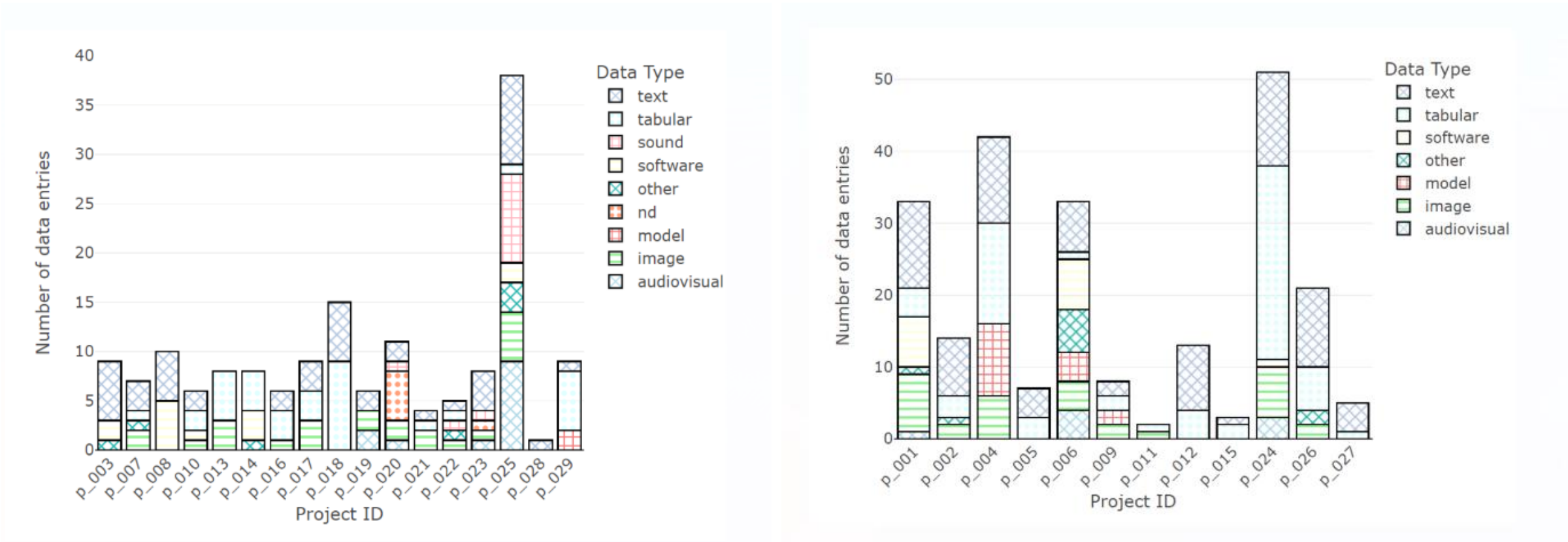

The data stewards then examined how data types are distributed across the broad subject areas used at UniBO to classify its 31 Departments. Each project was assigned to a subject area based on the Department of its Principal Investigator. The distribution of data types across subject areas is presented in Figure 5.

*Figure 5 Breakdown of data types chosen across subject areas.*

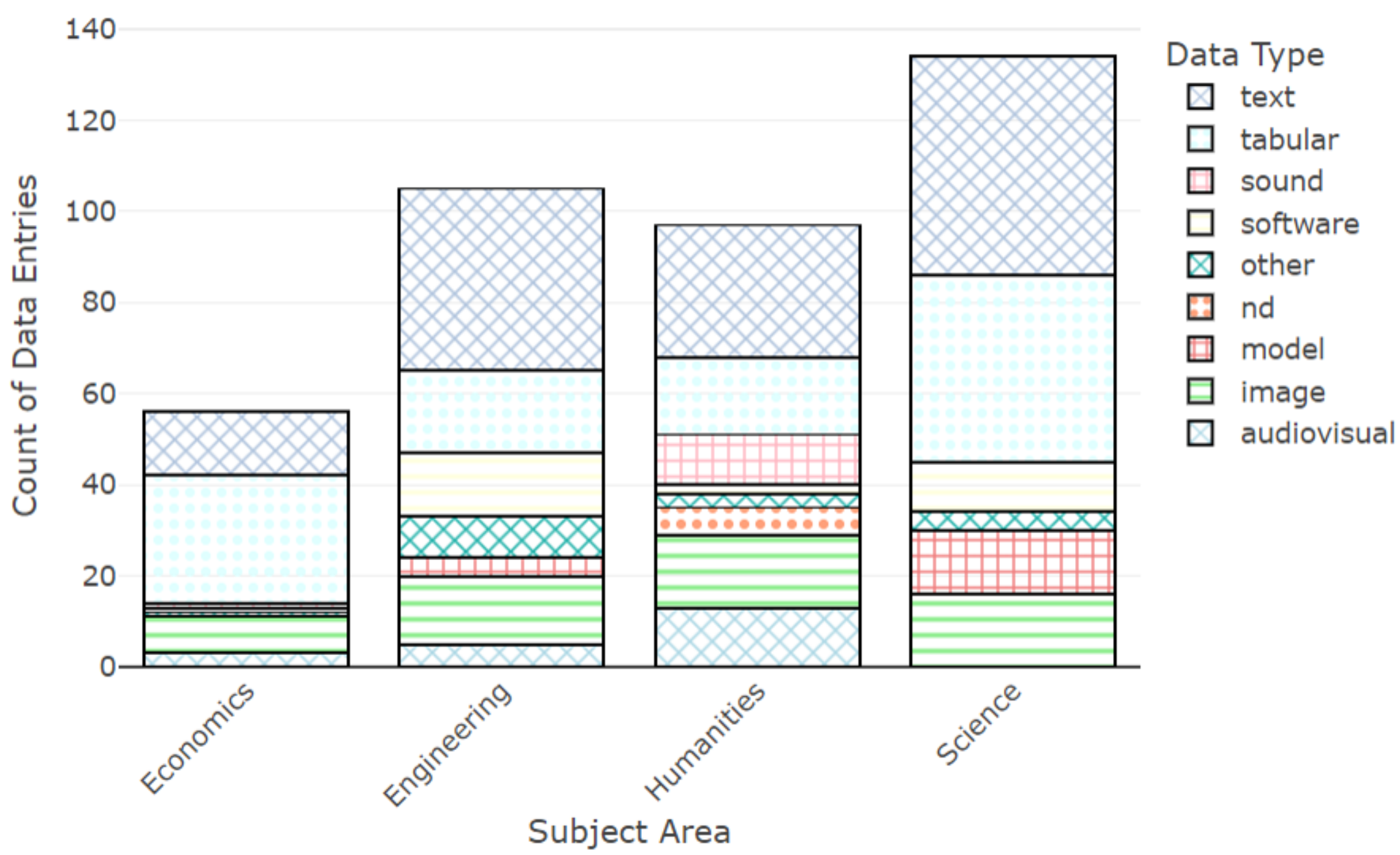

*Text* is the most common data type across all areas except Economics, where *Tabular* data prevails. Humanities disciplines make extensive use of *Sound* and *Audiovisual* data, particularly as raw materials for interviews. Engineering and Science projects more frequently produce *Software* and *Models*. *Image* data is widely used and evenly distributed across Engineering, Humanities, and Science. However, it is important to note that not all subject areas are represented in this study: for instance, there are no DMPs from the medical field.

The Humanities show a disproportionately high number of undetermined data types, largely due to one DMP that is notably vague and defers key details to a later stage.

### 1.2 How many DMPs include reused data, and what is the new-to-reused data ratio?

Out of the 29 DMPs analyzed, 12 describe reused data alongside newly generated data. Half of these (six) belong to the Humanities.

Across these DMPs, there are 2.33 new data entries for every reused one. However, reused data may be underreported in the DMPs.

### 1.3 What formats do researchers select for their data?

In most of the analyzed DMPs, the described entries are already linked to specific data formats, while only 43 out of the 392 total data entries have an undefined format. The most used data formats are shown in Table 3, and an overview of the broader range of formats is provided in Figure 6.

Some entries are associated with multiple data formats, reaching up to ten in certain cases, which indicates that format selection remains open. Choices are expected to be made as the project progresses, depending on workflow decisions such as tool selection, interoperability needs, or long-term preservation strategies.

*Table 3: Top 20 data format preferred by researchers*

| **Rank** | **Format** | **N of entries** |
|---|---|---|
| 1 | txt | 95 |
| 2 | csv | 78 |
| 3 | jpg | 33 |
| 4 | tiff | 29 |
| 5 | odt | 28 |
| 6 | xlsx | 25 |
| 7 | pdf | 24 |
| 8 | xls | 24 |
| 9 | png | 20 |
| 10 | rtf | 20 |
| 11 | mp4 | 19 |
| 12 | docx | 14 |
| 13 | mat | 12 |
| 14 | opj | 12 |
| 15 | doc | 11 |
| 16 | mp3 | 11 |
| 17 | ods | 11 |
| 18 | m | 9 |
| 19 | bag | 7 |
| 20 | xml | 7 |

Standard and open formats like .txt for text, .csv for tabular data, and .tiff/.jpg for images are commonly preferred. Widely used proprietary formats that have become standard, such as .docx and .xlsx, appear alongside open alternatives like .odt and .rtf.

*Figure 6 Word cloud showing the most cited data formats in the DMPs.*

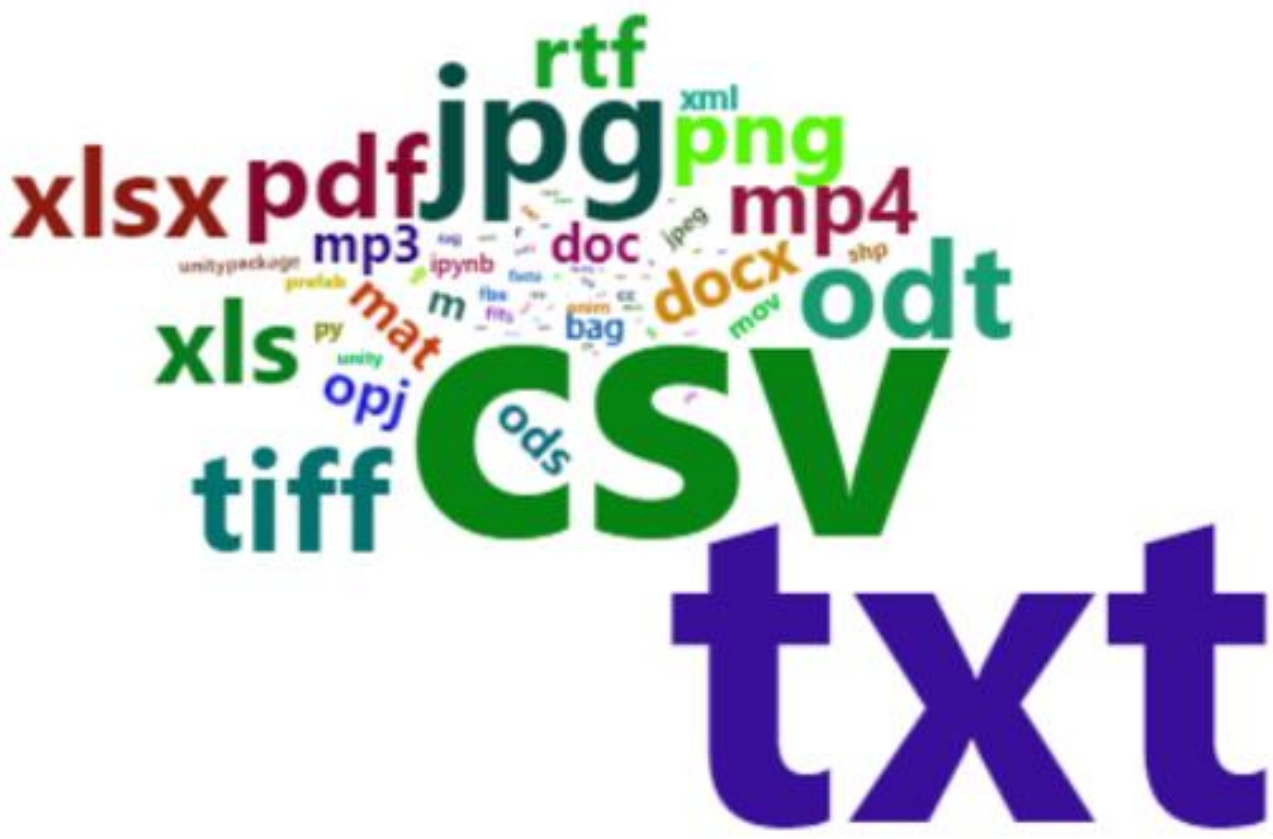

The data stewards note that the overall picture is reassuring. The prevalence of preservation-friendly formats likely reflects the influence of their support during the DMP drafting process.

## Research question 2. What recurring issues and patterns in data management could inform or improve data stewardship services?

### 2.1 How many projects handle personal data?

As presented in Figure 7, 15 out of 29 projects do not involve personal data processing at all. The absence of DMPs from the medical field may affect this result. Nonetheless, the data stewards identified 84 data entries requiring attention to personal data issues, and 10 entries where the use of personal data remains unclear.

*Figure 7 Need for personal data management for each project, evaluated per data entry.*

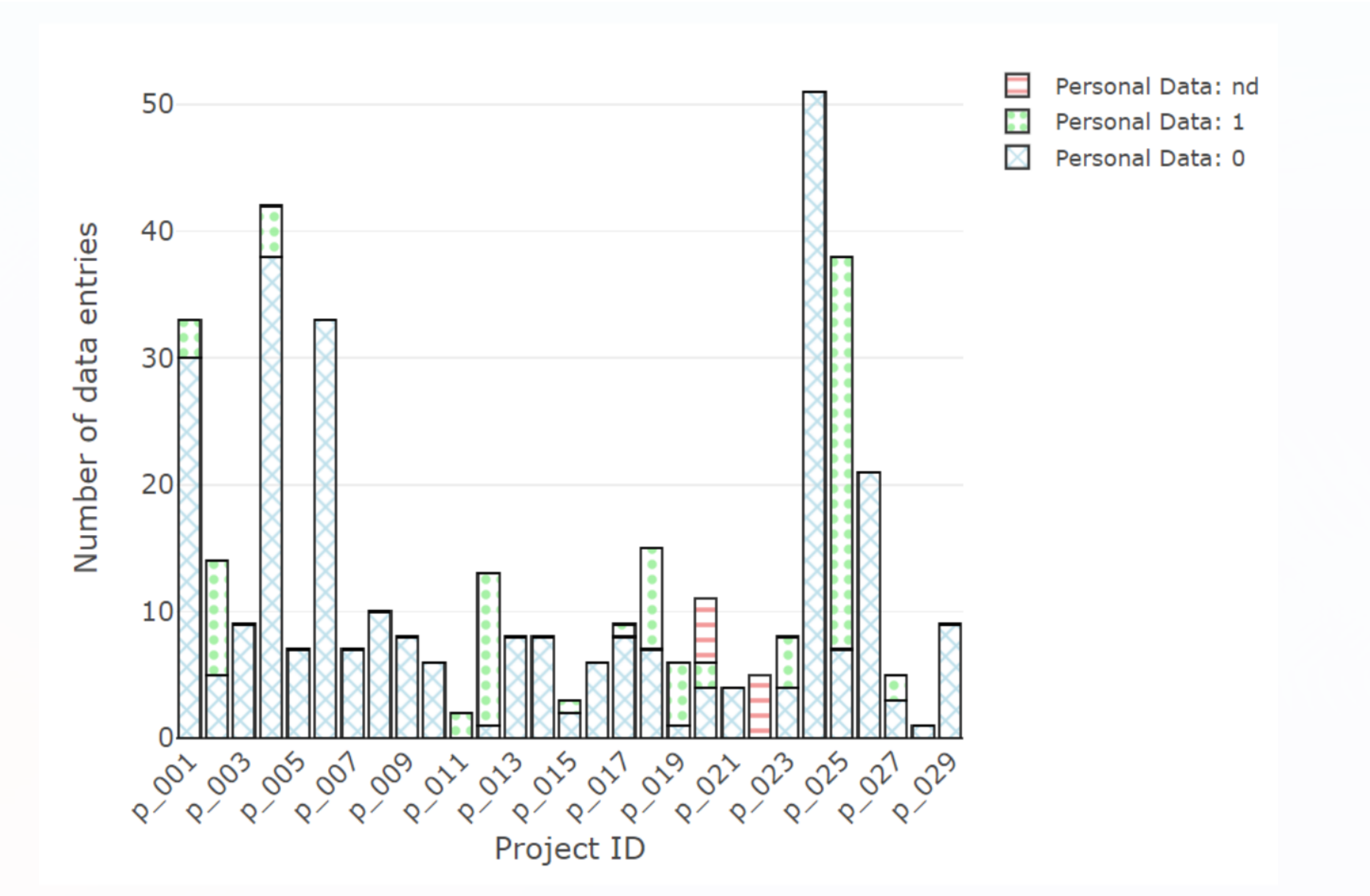

Exploring personal data use in EU-funded projects was particularly interesting to UniBo's data stewards because privacy regulations, such as GDPR and national laws, and related safeguarding measures, like encryption and anonymization, are a crucial and longstanding part of researchers' training at UniBO.

Figure 8 shows that in most cases (62), anonymization is the most common approach to ensure that data entries comprising personal data can be made public in compliance with privacy regulations. The analysis highlighted other strategies, though less frequently applied: in six cases researchers chose to apply pseudonymization to their data entries, in five cases they obtained consent for data publication. Interestingly, in 10 entries no protection strategy was indicated, while in one case it has still to be decided.

*Figure 8 Personal data management strategy per data entry.*

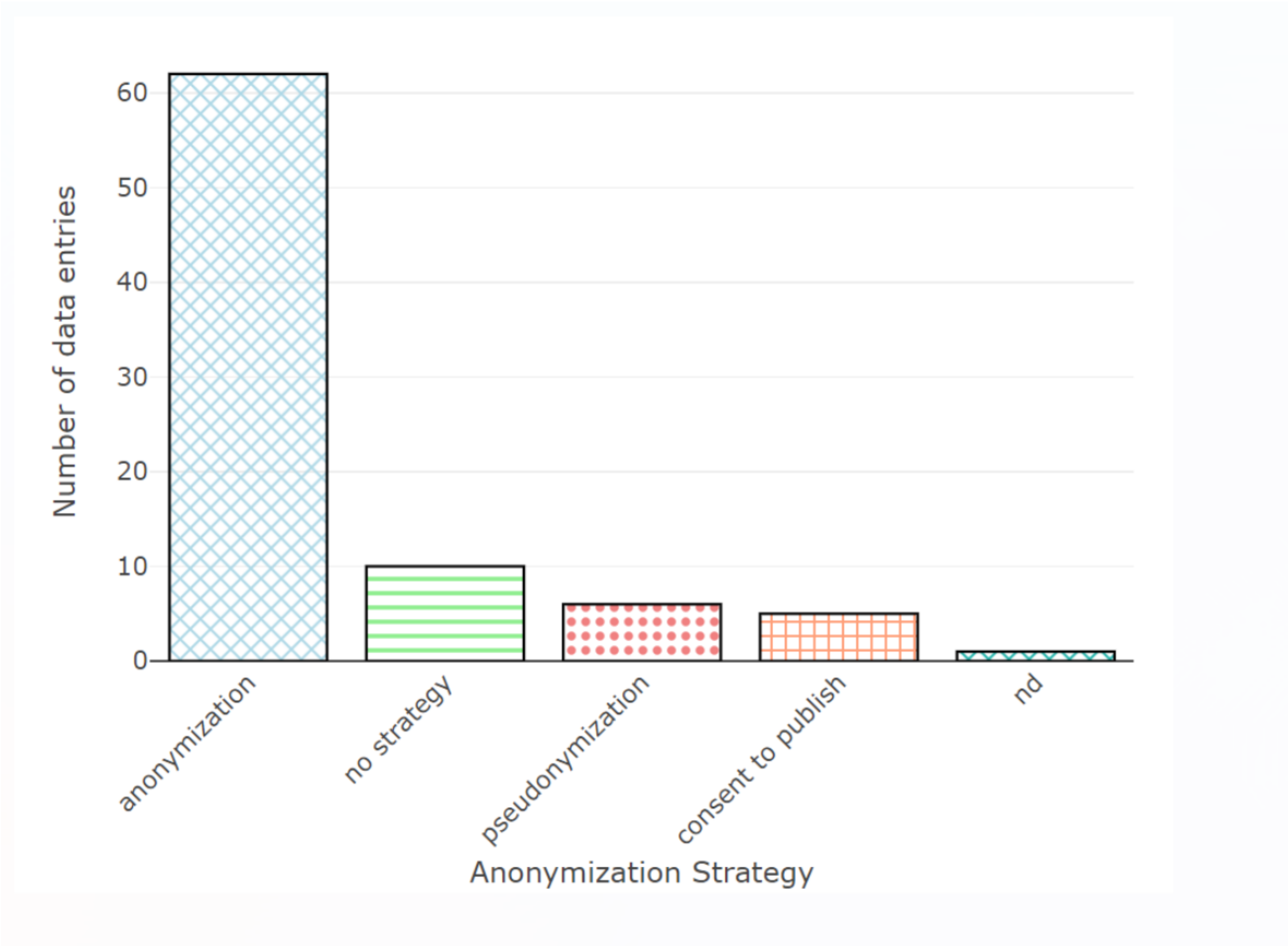

It is important to emphasize that pseudonymization is not considered best practice under GDPR, as pseudonymized data still fall under personal data regulations, even if the researcher lacks access to re-identification keys. Anonymization is therefore preferable, as only fully anonymized data are exempt from data protection laws.

However, in research contexts, pseudonymized data may offer scientific advantages. They can support reproducibility, result validation, and nuanced analysis over time, while still providing more protection than directly identifiable data. Although not optimal from a legal standpoint, pseudonymization can be a practical compromise, provided that robust technical and organizational measures are in place to minimize risks and ensure compliance. Under the right conditions, pseudonymized data may be deposited with controlled and limited access.

### 2.2 How many data entries remain closed, and why?

The data stewards examined data access levels based on specific data management strategies, focusing on cross-cutting concerns that affect data sharing and accessibility across domains, such as intellectual property rights (IPR), privacy, and ethics.

Access levels are defined as follows:

- Open: licensed under CC0, CC BY, CC BY-SA, or equivalents; for software: MIT, Apache, MPL.
- Accessible under conditions: licensed under CC BY-NC, CC BY-NC-SA, CC BY-ND, CC BY-NC-ND, or equivalents; for software: GPL, AGPL, JRL, AFPL.
- Controlled: access managed by the authors or a committee.
- Embargoed: temporarily restricted before becoming accessible.
- Unlicensed: freely available but lacking a formal open license.
- Unfiled: not deposited in a repository.
- Unknown: no information on accessibility or licensing.

Figure 9 visually presents the relation between the levels of access of different data entries with the major reasons for inaccessibility identified by researchers, while the frequency with which each aspect affects the level of access of the data entries is shown in table 4.

*Figure 9 Sankey diagram showing the influence of ethics, privacy, and IPR on the levels of access of the data entries.*

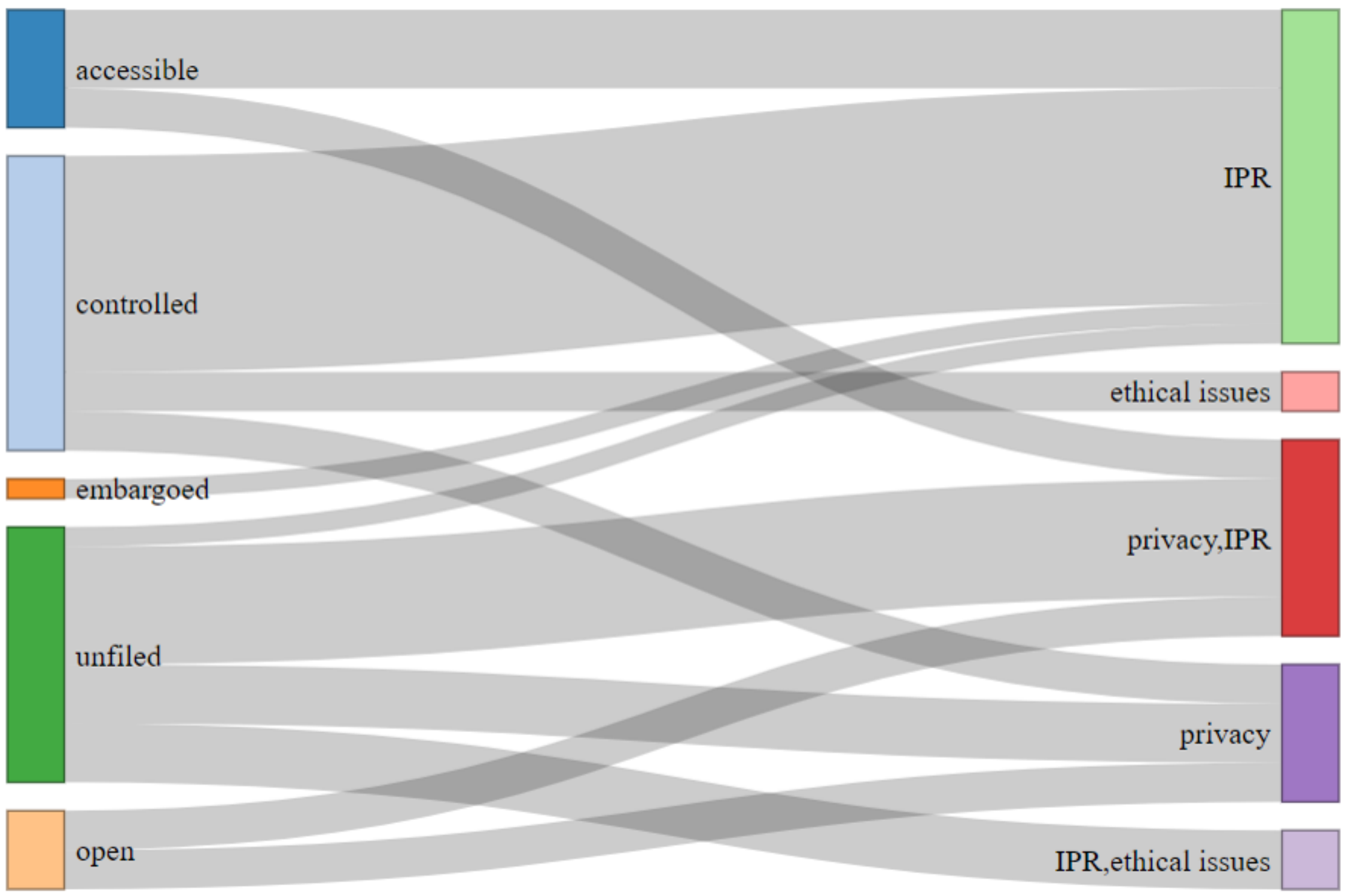

At month six, across all 29 DMPs, the plan is to release 15 data entries under controlled access. In 11 cases this is due to IPR concerns, in two to ethical issues, and in two to privacy. One data entry will be embargoed for IPR reasons. Four data entries will be accessible under conditions because of IPR, and two others for both privacy and IPR. Finally, four data entries will eventually be made available in open access after privacy issues (2) or both IPR and privacy issues are resolved (2).

Table 4: Frequency of ethics, privacy and IPR issues that may influence the levels of access

| Level of access | Reason of inaccessibility | N of entries |
|---|---|---|
| Open | Privacy | 2 |
| | Privacy, IPR | 2 |
| Accessible under conditions | Privacy | 4 |
| | Privacy, IPR | 2 |
| Controlled | Ethical issues | 2 |
| | IPR | 11 |
| | Privacy | 2 |
| Embargoed | IPR | 1 |
| Unfiled | IPR | 1 |
| | IPR, Ethical issues | 3 |
| | Privacy | 3 |
| | Privacy, IPR | 6 |

Some projects will not deposit data at all: 15 out of the total 392 data entries are marked as 'unfiled'. Three due to ethical and IPR concerns, six due to privacy and IPR, three due to privacy alone, and one due to IPR alone; for two further entries no explanation has been provided in the DMP.

Overall, significant variation in access strategies can be observed, despite the limited number of data points. This suggests that researchers are still navigating how to balance openness and security. While no clear pattern has emerged, such variation is to be expected in early research phases. It would be valuable to monitor how these practices evolve in EU-funded projects.

### 2.3 Is data size a common barrier?

In this study, most data entries (166) are estimated to be in the gigabyte range, 135 in megabytes, 11 in kilobytes, and only five in terabytes. For 75 entries, researchers did not provide size information in the DMP.

Data size can be a critical factor when selecting a repository for deposit: repositories vary in storage limits, both total volume and individual file size, which affects their suitability for large datasets. For example, Zenodo allows free deposits up to 50 GB per dataset.

Large datasets may also impact performance, slowing down retrieval, indexing, and processing. Additionally, costs can rise due to increased storage and bandwidth needs. Backup and recovery become more complex, and ensuring data integrity is more challenging, raising the risk of corruption. Regulatory compliance and security measures for large datasets are more resource-intensive, and usability can also be a concern, as accessing, visualizing and collaborating on large datasets often require specialized tools.

From the information provided in the DMPs analyzed here, data size does not appear to be a (potential) issue.

### 2.4 Which repositories are most chosen by researchers?

Table 5 shows that AMS Acta, UniBO's institutional repository, is the most used, with 109 entries, followed by Zenodo, with 94 entries. These platforms are preferred for their open access policies and alignment with funder mandates.

At month six, 82 data entries had no final repository decision. Of these, 16 were labeled 'nd' (no repository provided), while in all other cases the presence of multiple repositories for a single dataset indicated that some researchers had not yet finalized their choice.

Preferences of researchers indicating multiple repositories for a single data entry are not presented in Table 5, and they are distributed as follows: Zenodo and AMS Acta (52), AMS Acta and Zenodo (9), Zenodo and Bulgarian Portal for Open Science (8), Zenodo and ioChem-BD (6), OpenAgrar and Zenodo (4), Zenodo and MAST (2), European Nucleotide Archive and GenBank (1). Discipline-specific repositories also appear frequently, such as Chemotion for chemistry and Open Science Framework for general workflows. Platforms like FiVeR and OpenAgrar are chosen for agricultural data.

*Table 5: Repositories chosen by researchers for long-term preservation.*

| Chosen Repository | N of entries | URL |
|---|---|---|
| AMS Acta | 109 | https://amsacta.unibo.it/ |
| Zenodo | 94 | https://zenodo.org/ |
| Open Science Framework | 34 | https://osf.io/ |
| Chemotion | 12 | https://chemotion.net/ |
| FiVeR | 9 | https://fiver.ifvcns.rs/ |
| AMS Historica | 2 | https://historica.unibo.it/ |
| OpenAgrar | 2 | https://www.openagrar.de/content/index.xml |
| Springer website | 2 | https://www.springer.com/ |
| European Nucelotide Archive | 2 | https://www.ebi.ac.uk/ena/browser/home |

Overall, repository selection reflects a balance between institutional preferences, funder requirements, and disciplinary needs, with a clear inclination toward open and widely recognized platforms that support long-term accessibility.

### 2.5 How many researchers make their DMPs public?

Most researchers choose to make their DMPs publicly accessible, demonstrating a commitment to transparency: of the 29 projects analyzed, 20 DMPs are open to the public, allowing broader insight into data management strategies.

Seven DMPs remain restricted due to privacy, IPR, or ethical concerns. Two projects have yet to decide, possibly pending clarification on data protection or collaboration agreements.

This trend suggests increasing openness, while also highlighting the need to address legitimate concerns that may require restricted access.

### 2.6 How many projects adopt at least one standard?

Only seven out of 29 projects described in the DMPs have adopted standards, covering 15 total datasets. This low adoption rate suggests limited prioritization of interoperability, or that some fields lack established practices.

Standards, such as common formats, vocabularies, or protocols, are essential for interoperability, data quality, and reusability. The findings indicate that further outreach or training may be needed to promote broader adoption, especially in projects with complex or diverse datasets.

## Research question 3. Is there an interdisciplinary approach to data production at the University of Bologna?

To address this question, the data stewards considered only datasets newly produced by at least one UniBO researcher. Two theoretical premises are relevant:

- There is no universally accepted definition of interdisciplinarity.
- Interdisciplinary cases in the sample are limited, so the analysis is primarily qualitative.

The definition adopted is based on the Italian academic system of Scientific-Disciplinary Sectors (SSD), a formal taxonomy established by the Ministry for University and Research to classify research and teaching activities. Although a new SSD system was introduced in May 2024 (*D.M. 2 maggio 2024, n. 639, Determinazione dei gruppi scientifico-disciplinari*, no date), the study uses the previous version, as the research began before the update. The two systems are largely similar; equivalences can be found in Annex B of the decree (*D.M. 2 maggio 2024, n. 639, Determinazione dei gruppi scientifico-disciplinari - Allegato B*, no date).

SSDs are assigned to courses, professors, and researchers. Their distribution across departments at UniBO is uneven: some SSDs are concentrated in specific departments (e.g., SECS-P/08 Economics and Business Management in the Department of Management), while others appear across multiple departments (e.g., INF-01 Computer Science is represented at the Department of Computer Science and Engineering, at the Department of Philology and Italian Studies, at the Department of Legal Studies, and others). Departments also greatly vary in SSD diversity; for instance, the Department of Modern Languages, Literatures, and Cultures hosts 28 SSDs, while the Department of Chemistry hosts eight.

To increase granularity, the data stewards considered both the department and the SSD of the dataset creator(s) and project PI. SSDs offer a more precise disciplinary indicator, and since the analysis is local to UniBO, using the national SSD system is appropriate.

However, SSDs have limitations. Interdisciplinary fields like Digital Humanities lack dedicated SSDs and are instead split across existing ones (e.g., library studies, literary studies, computer science).

Despite these constraints, interdisciplinarity was tracked through:

1) Differences between the SSD of the dataset creator(s) and the PI.
2) Collaboration between creators with different SSDs.
3) The types of data produced in interdisciplinary contexts.
4) The prevalence of interdisciplinarity in collaborative vs. single-beneficiary projects.

Out of 392 data entries, 72 cases (about 18%) show a mismatch between the SSD of at least one dataset creator and that of the PI. These were classified as:

- Mild interdisciplinarity: SSDs differ but are adjacent or within the same disciplinary field (e.g., AGR/01 and AGR/02). This accounts for 39 cases (54% of interdisciplinary cases, 10% of the total).
- Strong interdisciplinarity: SSDs differ significantly and belong to unrelated fields. This includes 33 cases (46% of interdisciplinary cases, 8% of the total).

However, SSDs from different fields may still be closely related or commonly co-occur within departments, raising questions about whether these cases truly represent interdisciplinarity.

Moving to the second part of the analysis, 11 data entries involved collaboration between creators with different SSDs. Of these, eight were classified as strong interdisciplinarity.

The data stewards also examined whether interdisciplinarity influences the types of data produced. Among the 72 cases, textual data (28) and tabular data (22) were most common, in agreement with findings from RQ1.1.

Finally, the data stewards assessed whether interdisciplinarity is more frequent in collaborative or single-beneficiary projects. Single-beneficiary projects (e.g., ERC grants) often involve junior researchers alongside the beneficiary, who could indeed represent different SSDs. However, 67 out of 72 interdisciplinary cases occurred in collaborative projects, suggesting a predictable link between bigger research consortia and interdisciplinarity.

## Discussion and Conclusions

This paper outlines preliminary findings on data management practices adopted by researchers at UniBO. The analysis is based on a limited sample of DMPs from projects led by UniBO researchers and funded by European programs such as Horizon 2020 and Horizon Europe.

Since DMPs are mandatory for projects funded by the European Commission and central to their Open Science strategy, researchers are required to develop RDM procedures early on: this makes them not only more aware of these topics than the broader university population, but also frequent users of the university's data steward services.

Although the limited sample and researchers' familiarity with RDM may have influenced the analysis, it enabled the data stewards to identify common practices and challenges within an experienced cohort and to gather insights that can inform guidelines for the wider research community.

The analysis reveals that text and tabular data constitute the predominant data types managed by UniBO researchers, a trend that persists not only at the aggregate level but also within nearly all disciplinary domains. The predominance of these data types is reflected in the Data Management Plans (DMPs), in which .txt and .csv emerge as the most frequently referenced formats, both suitable for long-term preservation. Other data types, such as images and software, are also present, though less frequently. Although their distribution is uneven, these data types are represented across all disciplinary areas within the sample.

These findings indicate that data stewards should design their support strategies around shared good practices that are applicable across disciplinary boundaries. A greater overall impact could be achieved by prioritising cross-cutting guidelines and training initiatives that equip researchers with the fundamental skills required to manage commonly used data types and formats. It is also advisable to reuse and adapt resources developed by other institutions, thereby promoting efficiency and alignment with established standards.

Also, the inclusion in the DMPs of 'unconventional data' such as software and models indicates growing awareness of diverse research outputs. This suggests that a shift in perspective may be timely: rather than fitting all research outputs into a broad definition of data and including them into the RDM and DMP frameworks, the RDM community should consider a holistic view of research outputs (Chiesa and Sikder, 2024), emphasizing the importance of managing all analytical processes, tools, and knowledge structures (Houweling and Willighagen, 2023).

The results also show moderate awareness of dataset structuring and long-term preservation. Most projects, especially collaborative ones, manage multiple data types but tend to group same-type data into separate datasets, increasing the effort required for both structuring and preservation and often obscuring relationships between data created for a common research question.

Researchers often choose non-disciplinary repositories like Zenodo and AMS Acta for long-term preservation, which may reflect a lack of discipline-specific repositories. Fields like agronomy and genetics are exceptions to this trend and have well-established disciplinary repositories.

The use of general-purpose repositories can offer advantages when considering the increasingly interdisciplinary research landscape, as they are accessible to a broad range of researchers. While the present study did not allow direct links between RDM practices and interdisciplinary research, the data stewards still emphasize the importance of balancing discipline-specific approaches with the need for data interoperability to support mutual data sharing and understanding (Credi, Fariselli, *et al.*, 2024).

These observations suggest two complementary strategies for long-term preservation. According to European Commission guidelines (OpenAIRE, no date), discipline-specific repositories should be prioritized for specialized data, requiring homogeneous datasets. When no discipline-specific repository is available, the authors of this study believe that data stewards should advise organizing diverse data types within a single dataset for general-purpose repositories, an approach which enhances coherence, reflects data relationships, and supports reproducibility.

The analysis reveals that privacy compliance and intellectual property rights (IPR) significantly influence RDM strategies. Although the medical domain, typically reliant on personal data, is not represented in the sample, nearly half of the DMPs address privacy concerns, often through data anonymization strategies.

However, long-term preservation and sharing strategies for ethically sensitive, private, or IP-protected data are frequently undefined, indicating a cautious stance toward data reuse, an aspect further reflected in the limited planning for reusing pre-existing data. It is also important to remember that initial DMPs for EU-funded projects are mandatory within the first six months, which may constrain researchers' ability to assess complex reuse scenarios, leading to conservative and restrictive approaches.

These findings suggest that researchers benefit from an integrated RDM approach involving both data stewards and legal experts, supporting via this collaboration the integration of technical and legal considerations and fostering accountability (University of Bologna, no date). The prevalence of personal data in research projects highlights the need for involving privacy experts in guideline development and training, with a focus on promoting anonymization as a strategy to enhance data sharing and reuse.

The analysis was based on a limited sample of preliminary DMPs, which primarily reflect researchers' intentions rather than implemented practices. Nonetheless, the sample provided valuable insights into preferred strategies and informed institutional support and policy development. Table 1, with its list of variables for DMP content analysis, represents an important research contribution that could be useful in future DMP assessments carried out by others or by the same group. Applying the same methodology in future studies and conducting follow-up analyses would be beneficial to track the evolution of data management practices over time.

## Reproducibility

Sara Coppini, Bianca Gualandi, Giulia Caldoni, Mario Marino, Silvio Peroni, Francesca Masini 2024. Mapping Research Data at the University of Bologna: Protocol. Protocols.io https://dx.doi.org/10.17504/protocols.io.n2bvj87jpgk5/v2

The steps described in the protocol have been implemented in the code available here: Marino, M., Caldoni, G., Coppini, S., & Gualandi, B. (2024). Mapping Research Data at the University of Bologna: Code (Version 01). Zenodo. https://doi.org/10.5281/zenodo.14809078

The research results, organized in an open tabular file, are available here: Coppini, S., Caldoni, G., Gualandi, B., & Marino, M. (2024). Mapping Research Data at the University of Bologna: Dataset (Version 01) [Data set]. Zenodo. https://doi.org/10.5281/zenodo.14234555

The data management plan of the research is available here: Gualandi, B., Caldoni, G., Coppini, S., & Marino, M. (2024). Mapping research data at the University of Bologna: Data Management Plan (Version 1). Zenodo. https://doi.org/10.5281/zenodo.14385803

## Funding Information

This work has been partially funded by the European Union's Horizon Europe framework program under grant agreement No 101095129 (GraspOS Project).

## Competing Interests

The authors have no competing interests to declare.